\documentclass{article}
\usepackage{spconf,amsmath,graphicx,hyperref}
\usepackage{soul,color}
\usepackage{hyperref}

\def\etal{\emph{et al}.\ }
\title{ToS: A Team of Specialists ensemble framework for Stereo Sound Event Localization and Detection with distance estimation in Video}
\name{Davide Berghi, Philip J. B. Jackson\thanks{This research was funded by EPSRC-BBC Prosperity Partnership `{AI4ME}: Future personalised object-based media experiences delivered at scale anywhere' (EP/V038087/1). For the purpose of open access, the authors have applied a Creative Commons Attribution (CC BY) license to any Author Accepted Manuscript version arising. Data supporting this study is available from \url{https://zenodo.org/records/15559774}}}
\address{Centre for Vision, Speech, and Signal Processing (CVSSP)\\
        University of Surrey, Guildford, U.K.}
\begin{document}
%
\maketitle
\begin{abstract}

Sound event localization and detection with distance estimation (3D SELD) in video involves identifying active sound events at each time frame while estimating their spatial coordinates. This multimodal task requires joint reasoning across semantic, spatial, and temporal dimensions, a challenge that single models often struggle to address effectively. 
To tackle this, we introduce the Team of Specialists (ToS) ensemble framework, which integrates three complementary sub-networks: a spatio-linguistic model, a spatio-temporal model, and a tempo-linguistic model. 
Each sub-network specializes in a unique pair of dimensions, contributing distinct insights to the final prediction, akin to a collaborative team with diverse expertise. 
ToS has been benchmarked against state-of-the-art audio-visual models for 3D SELD on the DCASE2025 Task 3 Stereo SELD development set, consistently outperforming existing methods across key metrics.
Future work will extend this proof of concept by strengthening the specialists with appropriate tasks, training, and pre-training curricula.

\end{abstract}

\vspace{-1ex}
\section{Introduction}
\label{sec:intro}


Sound event localization and detection with distance estimation (3D SELD) \cite{Adavanne:2019:SELDnet,Krause:2024:seldDistance} 
is the joint problem of identifying active sound events from a predefined set of classes, track their temporal activity, and estimate their spatial coordinates, including direction or arrival (DOA) and distance. Therefore, it combines sound event detection (SED) \cite{Adavanne:2017:sed} and sound source localization (SSL) \cite{Adavanne:2018:DOA}, and can be tackled as a multimodal problem leveraging multichannel audio and video information. 
Since its introduction in the DCASE Challenge in 2019, SELD has evolved to address increasingly complex scenarios, including dynamic sound sources \cite{politis:2020:DCASE}, robustness to interference \cite{politis:2021:DCASE}, multimodal SELD in 360° videos \cite{Shimada2023STARSS23AA}, and distance estimation, i.e., 3D SELD \cite{Diaz-Guerra:2024:seldBaseline24}. 
The 2025 edition introduced stereo 3D SELD using frontal perspective video and stereo audio, replacing FOA and MIC audio formats, and 360$^\circ$ videos. 
This shift aligns better with mainstream media formats, but imposes new constraints: azimuth-only DOA estimation within [-90°, 90°] to avoid ambiguity in elevation and front-back discrimination. 
The task also includes predicting whether sound sources are onscreen or offscreen.

3D SELD inherently demands reasoning across three key dimensions: spatial cues for localization (space), temporal patterns for activity tracking (time), and semantic understanding for sound classification (semantics). Recent studies \cite{Berghi:2025:DCASE25techRep,Berghi:2025:dcaseworkshop} have enhanced semantic reasoning by integrating language-aligned encoders: CLAP \cite{wu:2023:clap_laion} for the audio modality, and replacing the commonly used ResNet50 \cite{He:2016:resnet} in audio-visual SELD pipelines \cite{Diaz-Guerra:2024:seldBaseline24,Shimada:2025:dcaseBaseline,Berghi:2024:ICASSP24,hong:2024:MVAnet} with OWL-ViT \cite{Minderer:2022:owl_vit}.
Despite these advancements, achieving robust joint reasoning across semantic, spatial, and temporal dimensions remains a significant challenge.
In this paper, we introduce the Team of Specialists (ToS), an ensemble framework specifically designed to navigate this multi-dimensional complexity. ToS comprises three dedicated sub-networks, each trained on the same 3D SELD task but optimized to focus on a distinct pair of dimensions. These include: a spatio-linguistic model, a spatio-temporal model, and a tempo-linguistic model. 
By leveraging complementary perspectives from each specialized network, ToS delivers enriched and more accurate final predictions. 
The outputs of the specialists are fused to produce robust SELD predictions. This modular design allows ToS to reason across all three dimensions without overburdening a single architecture.
We benchmark ToS against state-of-the-art audio-visual models for 3D SELD on the DCASE2025 Task~3 Stereo SELD development set. Our results show that ToS consistently outperforms existing methods across key metrics, demonstrating the effectiveness of a modular, ensemble-based approach to audio-visual SELD.

\begin{figure*}[bt]
\centerline{\includegraphics[width=165mm,clip,trim={0mm 15mm 0mm 0mm}]{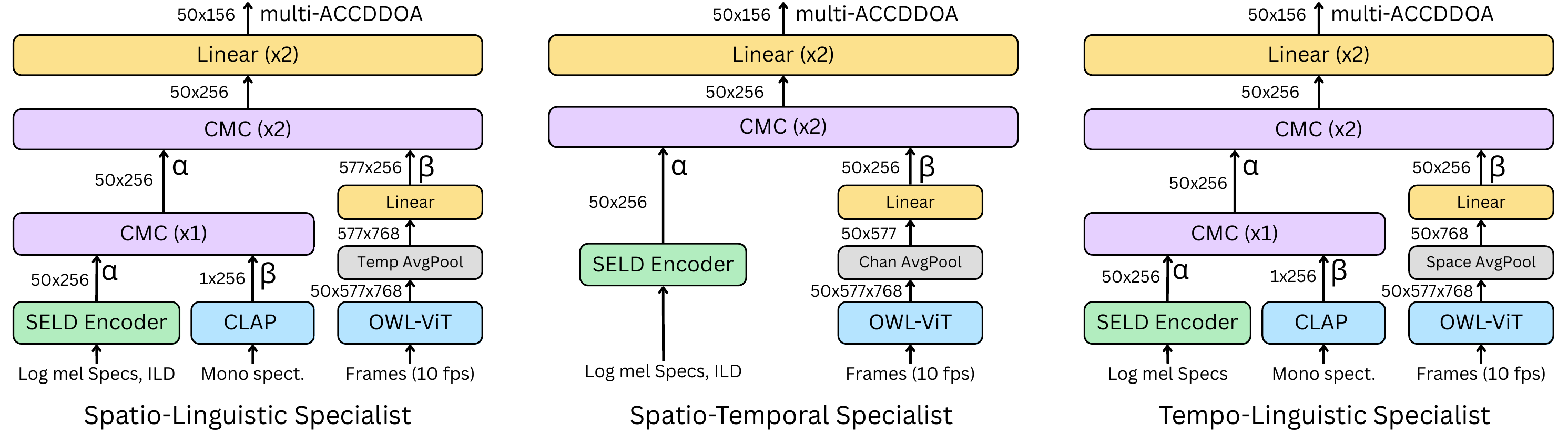}}
\vspace{-1ex}
\caption{Specialist audio-visual architectures: (left) Spatio-Linguistic, (center) Spatio-Temporal, (right) Tempo-Linguistic.}
\vspace{-2ex}
\label{Fig:tos}
\end{figure*}

The present work's contributions are:
(1)~to propose the Team-of-Specialists (ToS) framework for developing specialized audio-visual subnetworks to integrate spatial, temporal, and linguistic knowledge into a larger understanding system;
(2)~to evaluate both specialists and the ToS ensemble by benchmarking with the DCASE2025 Task3 challenge resources on 3D SELD using stereo audio and frontal video;
(3)~to demonstrate significant performance gains with respect to SOTA and alternate reference methods.
We now present the ToS framework, its component specialist models, describe our experiments and results, then conclude.

\vspace{-2ex}
\section{Team-of-Specialists (ToS) Framework}
\label{sec:tos}

To extract visual information relevant to SELD from video content, prior works \cite{Diaz-Guerra:2024:seldBaseline24,Shimada:2025:dcaseBaseline,Berghi:2024:ICASSP24,hong:2024:MVAnet} employed a pre-trained ResNet50 image encoder. The final layer of ResNet50 performs global average pooling over the resulting $7{\times}7$ spatial grid, producing a one-dimensional embedding vector. Processing each video frame, as in \cite{Berghi:2024:ICASSP24}, yields one embedding per frame. This pooling operation degrades spatial resolution, limiting the model's ability to leverage fine-grained spatial cues.
Other approaches \cite{Diaz-Guerra:2024:seldBaseline24,Shimada:2025:dcaseBaseline,hong:2024:MVAnet} retained the $7{\times}7$ spatial grid but averaged the visual embeddings across the channel dimension. We argue that this practice compromises the semantic richness extracted by ResNet50.
In our recent work \cite{Berghi:2025:dcaseworkshop}, we replaced ResNet50 with OWL-ViT \cite{Minderer:2022:owl_vit}, a language-aligned visual encoder trained for visual grounding tasks. 
OWL-ViT produces a fine-grained grid of visual embeddings.
We applied it to each video frame with temporal average pooling, preserving both spatial and semantic richness in the embeddings, albeit at the cost of temporal resolution.

Simultaneously managing time, space, and channel information in 
a large tensor is non-trivial to handle.
With ToS, the specialists leverage OWL-ViT's visual embedding, each pooling across a different dimension.
This uniformity of the feature embeddings is less pronounced in the audio modality. Spectrograms naturally encode time-frequency patterns, making them well-suited for enabling both temporal and semantic reasoning. Spatial cues can be derived from multichannel audio using features such as GCC-PHAT \cite{Knapp:gccphat:1976} or interchannel phase difference \cite{Nguyen:2021:SALSALiteAF} for microphone array (MIC) formats, or intensity vectors \cite{Perotin:2019:doaFeatures,cao:2020:EIN} for first-order Ambisonics (FOA) signals. These spatial features, combined with spectrograms, provide the network with rich and complementary information necessary to address SELD effectively.

\subsection{Spatio-Linguistic Specialist}
\label{subsec:sl}

Large language models (LLMs) \cite{Brown:2020:GPT}, vision-language models (VLMs) \cite{Radford:2021:clip}, and audio-language models (ALMs) \cite{Huang:2024:AudioGPT} have shown how language offers a powerful lens for semantic understanding and complex reasoning across media content.
The Spatio-Linguistic Specialist expands a conventional SELD architecture with CLAP \cite{wu:2023:clap_laion}, a language-aligned audio encoder, to enrich semantic representation. The core SELD encoder is a ResNet-Conformer \cite{Niu:2023:resnetConf,Wang:2023:ACS}, whose output embedding is fused with CLAP's via a Cross-Modal Conformer (CMC) \cite{Berghi:2025:dcaseworkshop}, a variant of the standard Conformer architecture \cite{Gulati2020ConformerCT}, adapted to cross-attend between two generic modalities ($\alpha$ and $\beta$).
On the visual branch, the specialist employs a pre-trained OWL-ViT \cite{Minderer:2022:owl_vit} applied to each video frame (10\,fps), and temporal pooling to support spatio-linguistic reasoning, emphasizing spatial and semantic dimensions. 
A final feed-forward module, consisting of two fully-connected layers, predicts multi-ACCDDOA vectors for up to $N{=}3$ tracks \cite{Krause:2024:seldDistance}, including on/off-screen activity as in the DCASE2025 Challenge baseline \cite{Shimada:2025:dcaseBaseline}. During training, the weights of CLAP and OWL-ViT are kept frozen. The Spatio-Linguistic Specialist is depicted in Fig.\,\ref{Fig:tos} (left).

While the Spatio-Linguistic Specialist architecture emphasizes spatial and semantic reasoning, temporal information is not entirely discarded. The audio inputs, comprising spectrograms and spatial features, are computed over time, allowing the model to capture temporal dynamics and produce frame-level predictions, as required by the SELD task.
This architecture builds upon the model proposed in our recent work \cite{Berghi:2025:dcaseworkshop}, with two key modifications. First, the convolutional backbone of the SELD encoder has been replaced with ResNet18, aligning with common practice in SELD literature \cite{Niu:2023:resnetConf,Wang:2023:ACS,hong:2024:MVAnet,Jiang:2024:AVseld,wang:2025:crossmodalDist}. Second, the embedding dimensionality has been reduced to 256, consistent with configurations used in \cite{hong:2024:MVAnet}, to ensure architectural compatibility and computational efficiency.

\vspace{-2ex}
\subsection{Spatio-Temporal Specialist}
\label{subsec:st}

The architecture of the Spatio-Temporal Specialist adapts that of the Spatio-Linguistic Specialist (described in Sec.\,\ref{subsec:sl}) making adaptations to emphasize temporal and spatial reasoning.
On the visual branch, OWL-ViT embeddings are extracted at each video frame and globally pooled across the channel dimension. This results in a sequence of spatial grids, one per frame, preserving spatial structure over time.
On the audio branch, only SELD-specific features are used. CLAP embeddings and the initial CMC are omitted as semantic and language reasoning are not this specialist's priority.
While this specialist is designed to prioritize spatial and temporal dimensions, audio spectrograms are included in the audio input feature set. This decision reflects the need to associate sound events with their DOA, which is key to SELD\@.
The Spatio-Temporal Specialist's architecture is in Fig.\,\ref{Fig:tos} (center).

\vspace{-1ex}
\subsection{Tempo-Linguistic Specialist}
\label{subsec:tl}

The architecture of the Tempo-Linguistic Specialist closely mirrors that of the Spatio-Linguistic, with modifications to emphasize temporal and semantic reasoning.
On the visual branch, each video frame is processed independently using OWL-ViT, and the resulting embeddings are averaged across the spatial dimension, yielding a temporally ordered sequence of semantic representations.
The audio branch includes both the SELD the CLAP encoders, whose outputs are fused via a CMC, as illustrated in Fig.\,\ref{Fig:tos} (right). To focus on time and semantics, spatial features are omitted from the SELD encoder's input. Nevertheless, some directional cues are implicitly captured through level differences in the input spectrograms, which retain spatial information to some extent.

All specialists, regardless of their primary focus, are trained to predict multi-ACCDDOA vectors for the 3D SELD task. While each specialist emphasizes a specific pair of dimensions, none entirely disregards the third. This balanced design ensures that all models retain the necessary capacity to solve the SELD problem, which inherently requires joint reasoning across space, time, and semantics.

\vspace{-1ex}
\subsection{Ensemble}
\label{subsec:ensemble}

To aggregate the predictions from the three specialists and produce the final ToS output, we adopt an ensemble strategy based on majority voting. A sound event is considered active at a given time frame if it is detected by at least two specialists, and their DOA predictions fall within a 20$^{\circ}$ angular threshold. The final DOA is then computed as the average of the contributing specialists' DOA estimates. Similarly, the distance is calculated as the mean of the predicted distances from those specialists.
For on/off-screen activity, if any one specialist predicts the event to be on-screen, it is considered to be within the camera's field of view (FoV).

\section{3D SELD Experiments}
\label{sec:Experiments}

\subsection{Dataset \& Input Features}

We benchmark our ToS framework on the DCASE 2025 Task3 Stereo SELD development set \cite{Shimada:2025:dcaseBaseline}. The stereo audio and frontal-perspective video are derived from STARSS23 \cite{Shimada2023STARSS23AA}, a MIC/FOA dataset with 360$^\circ$ video recordings. 
To obtain the stereo dataset, Shimada et al.\ \cite{Shimada:2025:dcaseBaseline} randomly select a viewing direction, crop a frontal 100$^\circ$ view from the video, 
and downmix aligned stereo signals from the FOA format, as in \cite{Mazzon2019FirstOA,Wilkins:2023:TwoVsFour,shimada:2025:SAVG}. 
The DCASE 2025 Task~3 Stereo SELD dataset has 30,000 5-second clips with a training-test split. 
Since the left and right audio channels are arithmetically derived from FOA signals (representing the spatial sound field at a point), we do not expect inter-channel time or phase differences.
So, we adopt inter-channel level difference (ILD) as the primary spatial feature for the SELD encoder, alongside log-mel spectrograms computed independently for each channel. ILD features are calculated as the ratio of the squared magnitudes of the short-time Fourier transforms (STFTs) of the left and right channels, then mapped into the log-mel domain: 
$\mathbf{ILD}(m,t) = \log\left[ \mathbf{H}_{\mathrm{mel}} \left( {|\mathbf{L}(f,t)|^2+ \epsilon} \right) / \left( {|\mathbf{R}(f,t)|^2 + \epsilon} \right) \right]$,
where $\mathbf{L}(f,t)$ and $\mathbf{R}(f,t)$ are the left and right channel STFTs, respectively, $\mathbf{H}_{\mathrm{mel}}$ the mel filter bank, $\epsilon$ a regularizing constant, and $m$ the mel-frequency index.
The full input feature set, i.e., two log-mel spectrograms plus ILD, are used as inputs to the SELD encoders of the Spatio-Linguistic and Spatio-Temporal specialists, whereas only log-mel spectrograms are given to the Tempo-Linguistic Specialist.
We did not include distance-related input features, such as stpACC \cite{Berghi:2025:distanceFeat}, 
which require extensive pre-training.
The input features are normalized for zero mean and unit standard deviation.
We trained our specialist models using a combination of real audio-visual data provided by the dataset and synthetic data generated using SpatialScaper \cite{Roman:2024:spatialScaper} and SELDVisualSynth \cite{roman2025generating}, following the methodology outlined in \cite{Berghi:2025:dcaseworkshop}. The synthetic dataset comprises 15,000 clips, each 5 seconds long, in line with the DCASE2025 Challenge baselines.

\vspace{-3pt}
\subsection{Implementation Details}

Audio spectrograms were computed using a short-time Fourier transform (STFT) with a 512-point Hann window and a 150-sample hop size. At 24 kHz sampling rate, this yields 800 temporal bins for each 5-second input clip. We use 64 mel bins for the log-mel input spectrograms.
The SELD encoder is a ResNet-Conformer, identical to the one used in \cite{Niu:2023:resnetConf,hong:2024:MVAnet}, with the exception of the pooling strategy. After each of the first three ResNet layers, frequency max pooling with a stride of 4 is applied, reducing the 64-bin frequency dimension to a single value. The final ResNet layer is followed by temporal average pooling with stride 4.
A second temporal average pooling layer with stride 4 is applied after the Conformer unit, reducing the 800 temporal bins to 50, which matches the label resolution, i.e., 10 labels per second with a 5-s input segment.
As in \cite{Berghi:2025:DCASE25techRep,Berghi:2025:dcaseworkshop}, we employed 4 Conformer layers in the SELD encoder, with 8 attention heads. 
For the CMCs, we used a single layer in the first CMC to fuse SELD and CLAP embeddings (for Spatio-Linguistic and Tempo-Linguistic specialists only), while two layers in the second CMC to fuse audio and visual embeddings.
The weights of the CLAP and OWL-ViT encoders were kept frozen, the rest of the specialists were trained on real and synthetic data for 100 epochs, using Adam optimizer, batch size 32, and a learning rate reduced by 5\% per epoch after the first 50 epochs.

Our evaluation uses the official metrics used in the DCASE 2025 Challenge\footnote{dcase.community/challenge2025/task-stereo-sound-event-localization-and-detection-in-regular-video-content\#evaluation}:
direction of arrival error (DOAE), relative distance error (RDE), on/off-screen  accuracy, and thresholded F1 score computed with on/off-screen predictions ($\mathrm{F_{\leq 20^{\circ}/1/on}}$) and without ($\mathrm{F_{\leq 20^{\circ}/1}}$).

\subsection{Results}
\label{subsec:results}

Table\,\ref{tab:tos} reports results achieved by each individual specialist, by pairwise ensembles of the specialists, and by the full ToS. The Spatio-Linguistic Specialist emerges as the top-performing specialist, likely due to its focus on arguably the two most challenging dimensions of SELD: semantic understanding and spatial localization. Temporal dynamics may be effectively captured by the SELD encoder itself, without explicit architectural emphasis.
The Spatio-Temporal Specialist shows a slightly lower F1 score compared to the others, which may be attributed to its reduced focus on semantic reasoning. Similarly, the Tempo-Linguistic Specialist exhibits weaker spatial resolution, reflecting its limited emphasis on spatial cues.
Pairwise ensembles improve DOAE and RDE but do not raise F1 compared to individual specialists. This is due to a different ensembling rule: with 3 specialists, an event is considered only when at least 2 specialists detect it, making ToS robust to false positives.
With only two specialists, a prediction is accepted if either detects it, increasing recall but reducing precision.
The ToS ensemble successfully combines the strengths of all three specialists, yielding a 3.3-point gain in $\mathrm{F_{\leq 20^{\circ}/1/on}}$ over the best-performing specialist, a 12\% relative increase. All evaluation metrics show improvement, with DOAE achieving a reduction of 22\% with respect to the Tempo-Linguistic Specialist.
This level of improvement is notable when compared to other ensembles reported in literature. In \cite{Berghi:2025:dcaseworkshop}, the ensemble model gave only a 4\% relative increase in F1 score over its best-performing component.

\begin{table}[bt]
\vspace{-6pt}
\caption{ 
Detection and localization performance of specialist networks, pairwise ensembles, and ToS ensemble: \mbox{ } $\mathrm{F_{\leq 20^{\circ}/1}}$ (F1) [\%], $\mathrm{F_{\leq 20^{\circ}/1/on}}$ (F1o) [\%], Direction Of Arrival Error (DOAE) [°], On/Off Accuracy (Acc) [\%]. 
SL: Spatio-Linguistic; ST: Spatio-Temporal; TL: Tempo-Linguistic.
}
\centerline{\begin{tabular}{c|c|c|c|c|c} \hline
\textbf{Specialist} & F1\,$\uparrow$ & F1o\,$\uparrow$ & DOAE\,$\downarrow$ & RDE\,$\downarrow$ & Acc\,$\uparrow$ \\ \hline
Spat-Ling & 36.9 & 27.4 & 18.9 & 32.5 & 79.2 \\
Spat-Temp & 34.5 & 26.2 & 19.5 & 35.7 & 79.9 \\
Temp-Ling & 35.1 & 26.6 & 21.3 & 33.4 & 79.9 \\ \hline

Ens(SL,ST) & 36.4 & 27.1 & 16.8 & 32.1 & 79.7 \\
Ens(SL,TL) & 36.5 & 27.3 & 17.2 & 31.5 & 79.5 \\
Ens(ST,TL) & 35.8 & 26.9 & 17.4 & 32.9 & 80.2 \\ \hline

ToS & \bf{40.2} & \bf{30.7} & \bf{16.7} & \bf{29.6} & \bf{80.3} \\
\hline
\end{tabular}
}
\label{tab:tos}
\vspace{-2mm}
\end{table}

\begin{table}[bt]
\vspace{-6pt}
\caption{
State-of-the-art and reference method comparison.
Metrics defined as Tab.\,\ref{tab:tos}.
AO: audio-only; AV: audio-visual.}
\centerline{\begin{tabular}{c|c|c|c|c|c} \hline
\textbf{Model} & F1\,$\uparrow$ & F1o\,$\uparrow$ & DOAE\,$\downarrow$ & RDE\,$\downarrow$ & Acc\,$\uparrow$ \\ \hline
DCASE AO 
& 22.8 & - & 24.5 & 41.0 & -  \\
DCASE AV 
& 26.8 & 20.0 & 23.8 & 40.0 & 80.0  \\ \hline
AV-Conf\cite{Berghi:2024:ICASSP24} & 33.9 & 25.7 & 19.1 & 34.1 & \bf{80.3} \\
Berghi\cite{Berghi:2025:dcaseworkshop} & 35.3 & 25.9 & 19.6 & 31.7 & 79.5 \\
MVANet\cite{hong:2024:MVAnet} & 34.7 & 24.4 & 19.5 & 32.8 & 79.6 \\
ToS (ours) & \bf{40.2} & \bf{30.7} & \bf{16.7} & \bf{29.6} & \bf{80.3} \\
\hline
\end{tabular}
}
\label{tab:benchmark}
\vspace{-2mm}
\end{table}

The proposed ToS framework is benchmarked with existing SOTA audio-visual methods: 
the audio-visual Conformer (AV-Conf) \cite{Berghi:2024:ICASSP24}, i.e., a Conformer module with concatenated audio and visual embeddings as input; MVANet by Hong \etal \cite{hong:2024:MVAnet} where a multi-stage attention network processes audio and visual data; and our recent model \cite{Berghi:2025:dcaseworkshop}, which is similar to the Spatial-Linguistic Specialist architecture, using a different convolutional backbone from ResNet18 for the SELD encoder, and an internal embedding dimension of 512. All methods are trained with the same training data, input features, and multi-ACCDDOA output representation for a fair comparison. 
Table\,\ref{tab:benchmark} compares the results, including the audio-only (AO) and audio-visual (AV) challenge baselines.

All SOTA methods substantially outperform the baseline systems in F1 score and localization. The three methods included in the comparison perform similarly overall, with Berghi \cite{Berghi:2025:dcaseworkshop} achieving the highest F1 score, and AV-Conf leading in DOAE and RDE. Among the three, AV-Conf reports a lower $\mathrm{F_{\leq 20^{\circ}/1}}$, but its higher on/off-screen accuracy elevates its $\mathrm{F_{\leq 20^{\circ}/1/on}}$ to second-best.
The proposed ToS ensemble outperforms all other SOTA methods, achieving an F1 score nearly 5 percentage points higher than Berghi and the lowest DOAE of 16.7$^{\circ}$. This performance gain, however, comes at the cost of increased computational overhead, as ToS aggregates predictions from three independent models.


\section{Conclusions}
\label{sec:conclusions}

We developed a novel machine-learning framework for audio-visual subnetworks to combine as a team of specialists for 3D sound event localization and detection.
To enable joint reasoning across semantic, spatial, and temporal dimensions, we trained spatio-linguistic, spatio-temporal, and tempo-linguistic networks, which we evaluated on the DCASE2025 Task~3 Stereo SELD dataset.
Although the spatio-temporal one was weaker and the spatio-linguistic stronger (surpassing all reference benchmarks), their ensemble increased the F1 scores more than 3~points with improvements across all metrics.
Further data augmentation and training are needed to test the ToS against state-of-the-art performance with extensive pre-training.
It would be interesting to investigate specialist pre-training curricula to fully exploit their complementarity. 

\vfill\pagebreak

{\ninept
\bibliographystyle{IEEEbib}
\bibliography{refs}

@string{icassp = "Proc. ICASSP"}

@string{interspeech = "Proc. Interspeech"}

@string{dcase = "Proc. DCASE"}

@string{eccv = "Proc. ECCV"}

@string{neurips = "Proc. NeurIPS"}

@string{cvpr = "Proc. CVPR"}

@inproceedings{politis:2021:DCASE,
    author = "{Politis {et al.}}, Archontis",
    title = "A Dataset of Dynamic Reverberant Sound Scenes with Directional Interferers for Sound Event Localization and Detection",
    booktitle = "DCASE Workshop",
    year = "2021",
    pages_ = "125--129",
    isbn = "978-84-09-36072-7",
    doi. = "10.5281/zenodo.5770113",
}

@inproceedings{politis:2020:DCASE,
  title = "A Dataset of Reverberant Spatial Sound Scenes with Moving Sources for Sound Event Localization and Detection",
  author = "Archontis {Politis et al.}",
  year = "2020",
  booktitle = "DCASE Workshop",
}

@ARTICLE{Wang:2023:ACS,
  author={{Wang {et al.}}, Qing},
  journal={IEEE/ACM TASLP}, 
  title={A Four-Stage Data Augmentation Approach to {ResNet-Conformer} Based Acoustic Modeling for Sound Event Localization and Detection}, 
  year={2023},
  volume={31},
  number={},
  pages={1251-1264},
  doi={10.1109/TASLP.2023.3256088}
}

@INPROCEEDINGS{Niu:2023:resnetConf,
  author_={Niu, Shutong and Du, Jun and Wang, Qing and Chai, Li and Wu, Huaxin and Nian, Zhaoxu and Sun, Lei and Fang, Yi and Pan, Jia and Lee, Chin-Hui},
  author={{Niu {et al.}}, Shutong},
  booktitle={ICASSP}, 
  title={An Experimental Study on Sound Event Localization and Detection Under Realistic Testing Conditions}, 
  year={2023},
  volume={},
  number={},
  pages={1-5},
}

@inproceedings{Huang:2024:AudioGPT,
author = {{Huang {et al.}}, Rongjie},
title = {{AudioGPT}: understanding and generating speech, music, sound, and talking head},
year = {2024},
booktitle = {AAAI Conf.\ on Artificial Intelligence},
}

@inproceedings{Diaz-Guerra:2024:seldBaseline24,
title = "Baseline models and evaluation of sound event localization and detection with distance estimation in {DCASE} 2024 {C}hallenge",
author = "{Diaz-Guerra {et al.}}, David",
year = "2024",
pages = "41--45",
booktitle = "DCASE Workshop",
}

@inproceedings{Gulati2020ConformerCT,
  title={Conformer: Convolution-augmented Transformer for Speech Recognition},
  author={Anmol {Gulati {et al.}}},
  year={2020},
  booktitle={Interspeech},
  pages={5036--5040}
}

@ARTICLE{Perotin:2019:doaFeatures,
  author_={Perotin, Lauréline and Serizel, Romain and Vincent, Emmanuel and Guérin, Alexandre},
  author={{Perotin {et al.}}, Lauréline},
  journal={IEEE Journal of Selected Topics in Signal Processing}, 
  title={{CRNN}-Based Multiple {DoA} Estimation Using Acoustic Intensity Features for Ambisonics Recordings}, 
  year={2019},
  volume={13},
  number={1},
  pages={22-33},
  doi={10.1109/JSTSP.2019.2900164}
}

@article{wang:2025:crossmodalDist,
    author_ = {Qing Wang and Ya Jiang and Hang Chen and Sabato Marco Siniscalchi and Jun Du and Jianqing Gao},
    author = {{Wang {et al.}}, Qing},
    title = {Cross-Modal Knowledge Distillation with Multi-Level Data Augmentation for Low-Resource Audio-Visual Sound Event Localization and Detection},
    journal   = {ArXiv},
    volume={2508.12334},
    year = {2025},
}

@INPROCEEDINGS{He:2016:resnet,
  author={{He {et al.}}, Kaiming},
  booktitle={IEEE Conf.\ on Comp.\ Vis.\ \& Pattern Recogn.\ (CVPR)}, 
  title={Deep Residual Learning for Image Recognition}, 
  year={2016},
  volume={},
  number={},
  doi={10.1109/CVPR.2016.90},
  pages={770-778},
}

@inproceedings{Adavanne:2018:DOA,
    title = {Direction of Arrival Estimation for Multiple Sound Sources Using Convolutional Recurrent Neural Network},
    author = "Sharath Adavanne and Archontis Politis and Tuomas Virtanen",
    author_ = {Adavanne {et al.}, Sharath},
    year = {2018},
    doi = {10.23919/EUSIPCO.2018.8553182},
    pages = {1462--1466},
    booktitle_ = {European Signal Processing Conference},
    booktitle = {EUSIPCO},
}

@InProceedings{cao:2020:EIN,
  title={Event-Independent Network for Polyphonic Sound Event Localization and Detection},
  author_={Cao, Yin and Iqbal, Turab and Kong, Qiuqiang and Zhong, Yue and Wang, Wenwu and Plumbley, Mark D},
  author={Cao {et al.}, Yin},
  booktitle_={Detection and Classification of Acoustic Scenes and Events Workshop},
  booktitle={DCASE Workshop},
  year={2020}
}

@INPROCEEDINGS{Jiang:2024:AVseld,
  author={{Jiang {et al.}}, Ya},
  booktitle={ICME}, 
  title={Exploring Audio-Visual Information Fusion for Sound Event Localization and Detection In Low-Resource Realistic Scenarios}, 
  year={2024},
  volume={},
  number={},
  pages={1-6},
  doi={10.1109/ICME57554.2024.10687782}
}

@inproceedings{Mazzon2019FirstOA,
  title={First Order Ambisonics Domain Spatial Augmentation for {DNN}-based Direction of Arrival Estimation},
  author_={Luca Mazzon and Yuma Koizumi and Masahiro Yasuda and Noboru Harada},
  author={Luca {Mazzon {et al.}}},
  booktitle_={Workshop on Detection and Classification of Acoustic Scenes and Events},
  booktitle={DCASE Workshop},
  year={2019},
}

@inproceedings{Berghi:2024:ICASSP24,
    author = {{Berghi {et al.}}, Davide},
    title = {Fusion of audio and visual embeddings for sound event localization and detection},
    booktitle = {ICASSP},
    year = {2024},
}

@article{roman2025generating,
  title={Generating Diverse Audio-Visual 360{$^{\circ}$} Soundscapes for Sound Event Localization and Detection},
  author_={Roman, Adrian S and Chang, Aiden and Meza, Gerardo and Roman, Iran R},
  author={{Roman {et al.}}, Adrian S},
  journal={ArXiv},
  year={2025},
  volume={abs/2504.02988}
}

@inproceedings{Berghi:2025:dcaseworkshop,
    author = "Berghi, Davide and Jackson, Philip J. B.",
    title = "Integrating Spatial and Semantic Embeddings for Stereo Sound Event Localization in Videos",
    booktitle = "DCASE Workshop",
    year = "2025",
}

@inproceedings{Brown:2020:GPT,
 author = {{Brown {et al.}}, Tom},
 booktitle = {NeurIPS},
 title = {Language Models are Few-Shot Learners},
 year = {2020}
}

@INPROCEEDINGS{wu:2023:clap_laion,
  author={{Wu {et al.}}, Yusong},
  booktitle={ICASSP}, 
  title={Large-Scale Contrastive Language-Audio Pretraining with Feature Fusion and Keyword-to-Caption Augmentation}, 
  year={2023},
}

@inproceedings{Radford:2021:clip,
  author       = {Alec {Radford {et al.}}},
  title        = {Learning Transferable Visual Models From Natural Language Supervision},
  booktitle    = {Int.\ Conf.\ on Machine Learning},
  year         = {2021},
}

@article{hong:2024:MVAnet,
  title={{MVANet}: {M}ulti-Stage Video Attention Network for Sound Event Localization and Detection with Source Distance Estimation},
  author={Hengyi {Hong {et al.}}},
  journal={ArXiv},
  year={2024},
  volume={abs/2411.14153}
}

@article{Berghi:2025:distanceFeat,
  title={Reverberation-based Features for Sound Event Localization and Detection with Distance Estimation},
  author={Davide Berghi and Philip J. B. Jackson},
  journal={ArXiv},
  year={2025},
  volume={abs/2504.08644},
}

@INPROCEEDINGS{Nguyen:2021:SALSALiteAF,
  author_={Tho Nguyen, Thi Ngoc and Jones, Douglas L. and Watcharasupat, Karn N. and Phan, Huy and Gan, Woon-Seng},
  author={{Tho Nguyen {et al.}}, Thi Ngoc},
  booktitle_={IEEE International Conference on Acoustics, Speech and Signal Processing}, 
  booktitle={ICASSP}, 
  title={{SALSA}-{L}ite: {A} Fast and Effective Feature for Polyphonic Sound Event Localization and Detection with Microphone Arrays}, 
  year={2022},
  volume={},
  number={},
  pages={716-720},
  doi={10.1109/ICASSP43922.2022.9746132}
}

@article{shimada:2025:SAVG,
  title={{SAVGBench: Benchmarking} Spatially Aligned Audio-Video Generation},
  author={{Shimada {et al.}}, Kazuki},
  author_={Shimada, Kazuki and Simon, Christian and Shibuya, Takashi and Takahashi, Shusuke and Mitsufuji, Yuri},
  journal={ArXiv},
  year={2025},
  volume={abs/2412.13462}
}

@inproceedings{Minderer:2022:owl_vit,
author = {{Minderer {et al.}}, Matthias},
title = {Simple Open-Vocabulary Object Detection},
year = {2022},
booktitle = {ECCV},
}

@INPROCEEDINGS{Krause:2024:seldDistance,
  author={{Krause {et al.}}, Daniel Aleksander},
  booktitle={EUSIPCO}, 
  title={Sound Event Detection and Localization with Distance Estimation}, 
  year={2024},
  volume={},
  number={},
  pages={286-290},
  doi={10.23919/EUSIPCO63174.2024.10715220}
}

@inproceedings{Adavanne:2017:sed,
    author_ = {Adavanne, Sharath and Pertil\"{a}, Pasi and Virtanen, Tuomas},
    author = {Adavanne {et al.}, Sharath},
    title = {Sound Event Detection Using Spatial Features and Convolutional Recurrent Neural Network},
    year = {2017},
    doi = {10.1109/ICASSP.2017.7952260},
    booktitle_ = {IEEE International Conference on Acoustics, Speech and Signal Processing},
    booktitle = {ICASSP},
    pages_ = {771–775},
    numpages = {5},
}

@article{Adavanne:2019:SELDnet,
  title={Sound Event Localization and Detection of Overlapping Sources Using Convolutional Recurrent Neural Networks},
  author={Sharath {Adavanne et al.}},
  journal={IEEE J.\ of Selected Topics in Sig.\ Proc.},
  year={2019},
  volume={13}, 
  pages={34-48}
}

@inproceedings{Berghi:2025:DCASE25techRep,
      title={Spatial and semantic embedding integration for stereo sound event localization and detection in regular videos}, 
      author={Davide Berghi and Philip J. B. Jackson},
      year={2025},
      booktitle={Technical Report of DCASE Challenge}, 
}

@INPROCEEDINGS{Roman:2024:spatialScaper,
  author={{Roman {et al.}}, Iran R.},
  booktitle={ICASSP}, 
  title={Spatial {S}caper: {A} Library to Simulate and Augment Soundscapes for Sound Event Localization and Detection in Realistic Rooms}, 
  year={2024},
  volume={},
  number={},
  doi={10.1109/ICASSP48485.2024.10446118}
}

@inproceedings{Shimada2023STARSS23AA,
  title={{STARSS}23: An Audio-Visual Dataset of Spatial Recordings of Real Scenes with Spatiotemporal Annotations of Sound Events},
  author={Kazuki {Shimada et al.}},
  booktitle={{NeurIPS}},
  year={2023},
}

@inproceedings{Shimada:2025:dcaseBaseline,
    author_ = "Shimada, Kazuki and Politis, Archontis and Roman, Iran R. and Sudarsanam, Parthasaarathy and Diaz-Guerra, David and Pandey, Ruchi and Uchida, Kengo and Koyama, Yuichiro and Takahashi, Naoya and Shibuya, Takashi and Takahashi, Shusuke and Virtanen, Tuomas and Mitsufuji, Yuki",
    author = "{Shimada {et al.}}, Kazuki",
    title = "STEREO SOUND EVENT LOCALIZATION AND DETECTION WITH
ONSCREEN/OFFSCREEN CLASSIFICATION",
    booktitle = "DCASE Workshop",
    year = "2025",
}

@ARTICLE{Knapp:gccphat:1976,
  author={Knapp, Charles and Carter, G. Clifford},
  journal={IEEE Transactions on Acoustics, Speech, and Signal Processing}, 
  title={The generalized correlation method for estimation of time delay}, 
  year={1976},
  volume={24},
  number={4},
  pages={320-327},
  doi={10.1109/TASSP.1976.1162830}
}

@inproceedings{Wilkins:2023:TwoVsFour,
    author_ = "Wilkins, Julia and Fuentes, Magdalena and Bondi, Luca and Ghaffarzadegan, Shabnam and Abavisani, Ali and Bello, Juan Pablo",
    author = "{Wilkins {et al.}}, Julia",
    title = "Two vs. Four-Channel Sound Event Localization and Detection",
    booktitle_ = "Proceedings of the 8th Detection and Classification of Acoustic Scenes and Events 2023 Workshop (DCASE2023)",
    booktitle = "DCASE Workshop",
    year = "2023",
    pages = "216--220",
}
}

\end{document}